\documentclass{llncs}
\usepackage{amssymb,amsmath}
\usepackage{todonotes}
\usepackage{stmaryrd}
\usepackage{multirow}
\usepackage{graphics}
\usepackage{amsmath,amssymb}
\usepackage[linesnumbered]{algorithm2e}

\usepackage{xcolor}

\newcommand{\N}{\mathbb{N}}

\newcommand{\R}{\mathbb{R}}
\renewcommand{\P}{\mathbf{P}}
\newcommand{\T}{\mathbf{T}}

\newcommand{\pnModel}{\mathcal{N}}
\newcommand{\pnPlaces}{\P}
\newcommand{\pnTransitions}{\T}
\newcommand{\pnInitialMarking}{M_0}
\newcommand{\pnPar}{\lambda}
\newcommand{\pnParameterSet}{\boldsymbol{\pnPar}}%In the preamble

\newcommand{\ccpn}{\N_{CC}}

\newcommand{\pnFiringIntervalFunction}{J_s}
\newcommand{\pnParDomain}{D_{\lambda}}
\newcommand{\pnParDomainOverApprox}{\bar{D}_{\lambda}}
\newcommand{\pnParMainDomain}{\mathbb{N}}
\newcommand{\pre}[1]{\ifmmode^{\bullet}\mspace{-2mu}#1\else$^{\bullet}\!\!#1$\fi}
\newcommand{\post}[1]{\ifmmode#1^{\bullet}\else$#1^{\bullet}$\fi}
\newcommand{\rd}[1]{\ifmmode^{\square}\mspace{-2mu}#1\else$^{\square}\!\!#1$\fi}
\newcommand{\inh}[1]{\mbox{$^\circ\! #1$}}
\newcommand{\pnLinCon}{\gamma}
\newcommand{\pnParValuation}{\nu}
\newcommand{\pnIntLeftPoint}[1]{ {}^{\uparrow} {#1}}
\newcommand{\pnIntRightPoint}[1]{ {#1}^{\downarrow}}

\newtheorem{Query}{Query}

\newcommand{\tpn}{TPN\xspace}

\newcommand{\ptpn}{P-TPN\xspace}

\newcommand{\gmec}{\textsc{Gmec}\xspace}
\newcommand{\gmecs}{\textsc{Gmec}s\xspace}

\newcommand{\newlyenabled}[1]{\ifmmode\uparrow\!enabled\left(#1\right)\else$\uparrow\!enabled\left(#1\right)$\fi}
\newcommand{\enabled}[1]{\ifmmode enabled\left(#1\right)\else$enabled\left(#1\right)$\fi}

\newcommand{\firable}[1]{\ifmmode firable\left(#1\right)\else$firable\left(#1\right)$\fi}

\newcommand{\st}{\; | \; }

\newcommand{\tctl}{TCTL\xspace}
\newcommand{\tpntctl}{\textit{TPN-TCTL}\xspace}
\newcommand{\ptpntctl}{\textit{PTPN-TCTL}\xspace}

\newcommand{\setZ}{\mathbb Z} 
\def\setN{{\rm I\!N}} % natural integers

\def\bbbq{{\mathchoice {\setbox0=\hbox{$\displaystyle\rm
Q$}\hbox{\raise
0.15\ht0\hbox to0pt{\kern0.4\wd0\vrule height0.8\ht0\hss}\box0}}
{\setbox0=\hbox{$\textstyle\rm Q$}\hbox{\raise
0.15\ht0\hbox to0pt{\kern0.4\wd0\vrule height0.8\ht0\hss}\box0}}
{\setbox0=\hbox{$\scriptstyle\rm Q$}\hbox{\raise
0.15\ht0\hbox to0pt{\kern0.4\wd0\vrule height0.7\ht0\hss}\box0}}
{\setbox0=\hbox{$\scriptscriptstyle\rm Q$}\hbox{\raise
0.15\ht0\hbox to0pt{\kern0.4\wd0\vrule height0.7\ht0\hss}\box0}}}}
 
\newcommand{\set}[1]{\left\{ {#1} \right\}}

\def\st{\; | \;}
\def\imply{\Rightarrow}

\def\G{{ \texttt{G} }}

\def\Gone{{ \texttt{G1} }}

\def\Lzero{{ \texttt{L0} }}
\def\Lone{{ \texttt{L1} }}
\def\PC{{ \texttt{PC} }}
\def\PCzero{{ \texttt{PC0} }}

\begin{document}
% AA: 1st version of title
%\title{
%Modeling and Analysis of Resilience Properties 
%using Parametric Time Petri Nets 
%for Biological Oscillatory Systems
%}

% MM: fix
%\title{
%Modeling and Analysis of 
%Resilience Properties in 
%Oscillatory Biological Systems using 
%Parametric Time Petri Nets
%}
\title{
Modeling of 
Resilience Properties in 
Oscillatory Biological Systems using 
Parametric Time Petri Nets\\
Supplementary Information
}
%\titlerunning{Hamiltonian Mechanics} 
\author{
Alexander Andreychenko\inst{1} 
\and Morgan Magnin\inst{2,3}
\and Katsumi Inoue\inst{2}
}
%
%\authorrunning{Ivar Ekeland et al.} % abbreviated author list (for running head)
%
%
\institute{Saarland University,\\
66123, Saarbrucken, Germany.
%\sr{to complete}\\
%\email{makedon@mosi.cs.uni-sb.de},\\
\and
National Institute of Informatics,\\
2-1-2, Hitotsubashi, Chiyoda-ku, Tokyo 101-8430, Japan.\\
\and
LUNAM Universit\'{e}, \'{E}cole Centrale de Nantes, IRCCyN UMR CNRS 6597\\
(Institut de Recherche en Communications et Cybern\'{e}tique de Nantes)\\
1 rue de la No\"{e} - B.P. 92101 - 44321 Nantes Cedex 3, France.
}

\maketitle

\begin{abstract}
Automated verification of living organism models allows us to gain 
previously unknown knowledge about underlying biological processes.
%\sr{In the same time it has been proved to be an effective method of
%model rectification}.
%AA: 1st approach in absract
%In this paper we show how parametric time Petri net-based framework
%can be applied to define the time behavior of
%biological oscillatory systems more precisely.
%We also demonstrate how changing of the environmental conditions 
%can be tackled in this framework to model-check the resilience properties
%of living organisms.
%We demonstrate the applicability of this technique using
%the simplified model of circadian clock where we
%aim at precise time characterization of the process
%and studying the influence of artif.\buildicial jet-lag.
% MM: fix
In this paper, we show the benefits to use
parametric time Petri nets in order to analyze 
precisely the 
%\sr{(quantitative)}
dynamic behavior 
of biological oscillatory systems.
In particular, we focus on the resilience properties of such systems.
This notion is crucial to understand the behavior of biological systems 
(e.g. the mammalian circadian rhythm) 
that are reactive and adaptive enough to endorse 
%various\sr{, but 
major
%,} 
changes in 
%its 
their
environment (e.g. jet-lags, day-night alternating work-time).
We formalize these properties through parametric TCTL 
and demonstrate how changes of the environmental conditions 
can be tackled to guarantee the resilience of living organisms. In particular, we are able to discuss the influence of various perturbations, e.g. artificial jet-lag or components knock-out, with regard to quantitative delays. This analysis is crucial when it comes to model elicitation for dynamic biological systems. We demonstrate the applicability of this technique using 
a simplified model of circadian clock.

\keywords{parametric time Petri net, resilience, biological oscillators, model checking}
\end{abstract}

\begin{appendix}
\section{Logical characterization}
\label{app:logical_characterization}
\subsection{Notations}

%The sets $\setN$, $\mathbb{Q}^+$ and $\mathbb{R}^+$ are respectively the sets of natural, non-negative rational and non-negative real numbers. An interval I of $\mathbb{R}^+$ is a $\mathbb{Q}^+$-interval iff its left endpoint belongs to $\mathbb{Q}^+$ and its right endpoint belongs to $\mathbb{Q}^+ \cup \{\infty\}$. We set $I^{\downarrow} = \{ x| x \le y$ for some $y \in I \}$, the \emph{downward closure} of $I$ and $I^{\uparrow} = \{ x| x \ge y$ for some $y \in I \}$, the \emph{upward closure} of $I$. We denote by $\mathcal{I}(\mathbb{Q}^+)$ the set of $\mathbb{Q}^+$-intervals of $\mathbb{R}^+$.
The sets $\setN$
and $\mathbb{R}^+$ are respectively the sets
of natural
non-negative real numbers.
An interval I
of $\mathbb{R}^+$ is a $\mathbb{N}$-interval iff its left endpoint belongs to
$\mathbb{N}$ and its right endpoint belongs to $\mathbb{N} \cup \{\infty\}$.
We set $I^{\downarrow} = \{ x \vert x \in \mathbb{R^+},  x \le y$ 
for some $y \in I \}$, 
the \emph{downward closure} of $I$ and 
$I^{\uparrow} = \{ x \vert x \in \mathbb{R^+},  x \ge y$ 
for some $y \in I \}$, the \emph{upward closure} of $I$. 
We denote by
$\mathcal{I}(\mathbb{N})$ the set of 
$\mathbb{N}$-intervals of $\mathbb{R}^+$.
%\todo{clarify the nature of $x$. Is it $x \in R$?}
	
	%\sr{Citations of couple of TCTL success stories?}

	\subsection{Parametric time Petri nets with read and logical inhibitor arcs}
	We consider the model-checking problem of parametric time Petri net 
	%with read and \so{logical} \todo{read arcs being always logical, I recommend to put the 
	%logical word only before inhibitor} inhibitor arcs models (PRITPNs) \todo{express that the 
	with read and logical inhibitor arcs models.
	%\todo{express that the 
	%models you will use in the following sections produce only bounded nets, 
	%meaning no additional expressivity for read and logical arcs + citation}. 
	This class of models allows us to use
	parametric temporal bounds for transitions.
	Therefore, the model-checking procedure addresses 
	the model verification versus the given property 
	together with the parameter synthesis problem.
	Here we use models that produce only bounded nets,
	so adding the read and logical inhibition arcs does 
	not add the expressivity
	to parametric time Petri Nets formalism~\cite{tpn_expressive_power_2013}.
	A parametric time Petri net with read and logical inhibitor arcs 
	(\ptpn)
	is a tuple 
	$\pnModel = (
								\pnPlaces, 
								\pnTransitions,
								\pnParameterSet,
								\pre{(.)}, \post{(.)}, \rd{(.)}, \inh{(.)}, \allowbreak 
								\pnInitialMarking, \allowbreak 
								\pnFiringIntervalFunction, \allowbreak 
								\pnParDomain
							 )$,
	%TODO here we can save space if we need compressed definition of PN
	%where $\pnPlaces, \pnTransitions$ and $\pnParameterSet$ 
	%are the sets of places, transitions	and parameters correspondingly, 
	%$\pre{(.)}, \post{(.)}, \inh{(.)},$
	where:
	\begin{itemize}
	\item $\pnPlaces=\{p_1,p_2,\ldots,p_m\}$ is a non-empty finite set of \emph{places},
	\item $\pnTransitions=\{t_1,t_2,\ldots,t_n\}$ is a non-empty finite set 
				of \emph{transitions},
	\item $\pnParameterSet=\{\pnPar_1,\pnPar_2,\ldots,\pnPar_l\}$ is a finite set 
				of non-negative natural \emph{parameters},
	\item $\pre{(.)} \in \left( \N^\pnPlaces \right)^\pnTransitions$ is the 
				\emph{backward incidence function},
	\item $\post{(.)} \in \left( \N^\pnPlaces \right)^\pnTransitions$ is the 
				\emph{forward incidence function},
	\item $\rd{(.)} \in \left( \N^\pnPlaces \right)^\pnTransitions$ is the
				\emph{read function},
	\item $\inh{(.)} \in \left( \N^\pnPlaces \right)^\pnTransitions$ 
				is the \emph{inhibition function},
	\item $\pnInitialMarking \in \N^P$ is the \emph{initial marking} of the net,
	\item $\pnFiringIntervalFunction \in 
				\left( \mathcal{J(\pnParameterSet)} \right)^\pnTransitions $
				is the function that associates a \emph{parametric firing interval}	
				to each transition,
	\item $\pnParDomain \subseteq \pnParMainDomain^{\pnParameterSet}$
				%\todo{isn't $D_p$ confusing notation?}
				is a convex polyhedron that is the \emph{domain of the parameters}.
	\end{itemize}
	The net $\pnModel$ is parametrized 
	with a set of temporal parameters $\pnParameterSet$ 
%	of non-negative integer
	%\todo{right now it is defined over integers not over rationals}
%	temporal parameters
	%These parameters are considered as constant variables in the	semantics.
	together with linear constraints that define the domain $\pnParDomain$.
	Linear constraints are given in the form 
	%\limits
	$\pnLinCon =  \sum_{i=0}^{l} a_i \pnPar_i \sim b$, 
	%where coefficients $a_i, b \in \pnParMainDomain$, $i=\overline{1,l}$
	%where coefficients $a_i, b \in \R$, $i=\overline{1,l}$
	where coefficients $a_i, b \in \R$, $i \in \{1,\ldots,l \}$
	%\todo{real coefficients possible?}
	and relation $\sim \in \{<, >, \leq, \geq, =\}$.
%	\todo{nature of $\lambda$. Is it rational?}
	We select the natural subset $\pnParDomain \subseteq \pnParDomainOverApprox$ 
	from the set $\pnParDomainOverApprox \subset {\R}^{\pnParameterSet}$
	defined by constrains $\gamma$.
%	\todo{say that it must be guaranteed that all minimum bounds 
%	of firing intervals must be lower or equal to maximum bounds?}
	A valuation of the parameters is a function 
	$\pnParValuation: \pnParameterSet \mapsto \N_0$
	that assigns the value to each parameter,
	i.e. 	
	$\left[ \pnParValuation(\pnPar_1), \ldots, \pnParValuation(\pnPar_l) \right]^T \in \pnParDomain$.
	We define a parametric time interval as a 
	function $\pnFiringIntervalFunction: \pnParDomain \mapsto \mathcal{I}(\N)$ 
	%\todo{no def of $\mathcal{I}(\N_0)$ is given, maybe not a problem}
	that associates an integer interval to each parameter valuation
	($\mathcal{I}(\N)$ denotes the set of $\N$-intervals).
	%TODO uncomment if more detailed definitions are needed
	%	The set of parametric time intervals over	Par is denoted by J(Par). 
	%	As for numerical interval, J can be split into two
	%	functions ? J and J ? that are respectively the minimum bound and the maximum
	%	bound. They can be both represented by a linear constraint over the parameters.
%	\todo{Here the notation of $I$, $I_s$, $J$, $J_s$ shall be checked}
	
	A \emph{marking} $M$ of the net is an element of $\mathbb{N}^P$ such that
	$p \in P$
	the number of tokens is $M(p)$.
	A transition $t$ is said to be \emph{enabled} 
	by the marking $M$ 
	if $\left[
			\left( M \geq	\pre{t} \right) 
			\wedge 
			\left( M \geq	\rd{t} \right)
			\wedge 
			\left( M <	\inh{t} \right)
			\right]$,
	%\todo{and $M < \inh{t}$?}
	%(\emph{i.e.} if the number of tokens in $M$ in each input place of
	%$t$ is greater or equal to the value on the arc from this place to the
	%transition).
	%We denote it by $t \in enabled(M)$.
	and denoted by $t \in enabled(M)$.
	%A transition $t$ is said to be \emph{inhibited} 
	%by the marking $M$ 
	%TODO detailed explanation
	%if all the places of one of its inhibitor hyperarcs 
	%are marked with a number of tokens at least 
	%equal to the weight of the inhibitor hyperarc between these places and $t$: 
	%$\exists i \leq I(t), \inh{t^{i}} \leq M$.
	%if $\inh{t} \leq M$
	%and denoted by $t \in inhibited(M)$.
	%TODO detailed explanation
	%We denote it by $t \in inhibited(M)$. 
	%Practically, inhibitor hyperarcs are used 
	% to stop the elapsing of time for some transitions: 
	% a branch inhibitor hyperarc between some places $p_1,\ldots, p_k$ 
	% and a transition $t$ means that 
	% the stopwatch associated to $t$ is stopped 
	% as long as all the places $p_1,\ldots, p_k$ are marked.
%	A transition $t$ is said to be \emph{active} 
%	in the marking $M$ if it is enabled and not inhibited by $M$.
	
	We define the semantics of a \ptpn $\pnModel$ 
	via the semantics of \tpn
	%\todo{any citation here?}
	$\llbracket \pnModel \rrbracket_\pnParValuation$
	by assuming the certain
	valuation of parameters 
	$\pnParValuation \in \pnParDomain$ such 
	that 
	$ \llbracket \pnModel \rrbracket_\pnParValuation =
		%\left(
		(
			\pnPlaces, 
			\pnTransitions,
			\pnParameterSet, \allowbreak 
			\pre{(.)},  \allowbreak 
			\post{(.)},  \allowbreak 
			\rd{(.)},  \allowbreak 
			\inh{(.)},  \allowbreak 
			\pnInitialMarking,  \allowbreak 
			I_s
		%\right)
		)
	$,
	where 
	$ \forall t \in \pnTransitions, I_s(t) = \pnFiringIntervalFunction(t)(\pnParValuation) $
	%.
	%as follows:
	and
	\begin{itemize}
	\item	a transition $t$ is firable if it has been enabled 
				%and not inhibited
				for at least $\pnIntLeftPoint{I_s}(t)$
				time units,
	\item a transition $t_k$ is said to be \emph{newly} enabled 
				(denoted by $\uparrow enabled(t_k, M , t_i)$)
				by the firing of the transition $t_i$ 
				from the marking $M$
				if the transition is enabled by the new marking 
				$M^{'} = M - \pre{t_i} + \post{t_i}$ 
				but was not by
				%$M - \pre{t_i}$, 
				$M$.
				%where $M$ is the marking of the net before the firing
				Formally,
				\begin{equation*}
				\begin{array}{l}
				\uparrow enabled(t_k, M , t_i) =
				\left[
					\left( \pre{t_k} \leq M^{'} \right) 
					\wedge 
					\left( \rd{t_k} \leq M^{'} \right)
					\wedge 
					\left( \inh{t_k} > M^{'} \right)
				\right]
				\wedge\\
				\left[ 
					\left( t_k = t_i \right) 
					\vee
					\left( \pre{t_k} > M \right) 
					\vee
					\left( \rd{t_k} > M \right) 
					\vee
					\left( \inh{t_k} \leq M \right) 
				\right].
				\end{array}
				\end{equation*}
				The set of transitions newly enabled by firing 
				the transition $t_i$ from the marking $M$
				is denoted by $\uparrow enabled(M, t_i)$,
	\item a state of \tpn is given by the pair $q=(M,I)$ 
				where $M$ is a marking 
				and 
				$I \in (\mathcal{I}(\N))^\pnTransitions$
				is an \emph{interval} function
				%TODO more detailed explanation
%				Function $I \in (\mathcal{I}(\mathbb{N}))^T$
				that
				associates a  temporal interval  with
				every transition enabled at $M$.
%				\todo{does $I$ changes in time? Or can it be also marking dependent?}
%				\todo{check it}
	\end{itemize}
	The semantics of a \tpn
	$\llbracket \pnModel \rrbracket_\pnParValuation$
	can be defined as a time transition system~\cite{larsen1995model}
	$\mathcal{S}_{\llbracket \pnModel \rrbracket_\pnParValuation } = (Q, q_0, \rightarrow)$,
	where two kinds of transitions are possible:
	\emph{time} transitions (when time elapses)
	and 
	\emph{discrete} transitions (when a transition of the net is fired),
	where:
	%\begin{definition}[Semantics of a RITPN]
	%\label{def:sem-dense-tpns}
	%The semantics of a RIPTPN $\pnModel$ is defined as a Timed Transition system 
	%$\mathcal{S}_{\mathcal{N}} = (Q, q_0, T, \rightarrow)$ such that:
	\begin{itemize} 
	\item $Q = \N^\pnPlaces \times \mathcal{I}(\N)^\pnTransitions$,
	\item $q_0 = \left( \pnInitialMarking, I_s \right)$,
	\item $\rightarrow \in Q \times (\pnTransitions \cup \N) \times Q$
	is the transition relation including a time transition relation 
	and a discrete transition relation.
	The time transition relation is defined $\forall d \in \N$ as:
	%\begin{itemize}
	%\item let $q=(M,\nu) \in Q$ and $q'=(M,\nu') \in Q$ be two states of the net, the 
	%continuous time transition relation is defined $\forall d \in \mathbb{R}^+$ by:
	\begin{equation*}
	\begin{array}{l}
	(M, I) \stackrel{d}{\longrightarrow} (M, I')
	\text{ iff	}
	\forall t_i \in \pnTransitions,\\
	\left\{
		\begin{array}{l}
			I'(t_i) = 
			\left\{
			\begin{array}{l}
			%\left( \pnIntLeftPoint{I'(t_i)}, \pnIntRightPoint{{I'(t_i)}} \right), \\
			( \pnIntLeftPoint{I'(t_i)}, \pnIntRightPoint{{I'(t_i)}} ), \\
			\quad  \pnIntLeftPoint{I'(t_i)} = \max(0, \pnIntLeftPoint{I(t_i)}-d),
			\pnIntRightPoint{{I'(t_i)}} = \pnIntRightPoint{{I(t_i)}}-d,\\
			\quad \text{ if } t \in enabled(M),\\
			I(t_i), \text{ otherwise },
			\end{array}
			\right.\\
			M \geq \pre{t_i} \Rightarrow \pnIntRightPoint{{I(t_i)}} \geq 0
	 	\end{array} 
	 	\right.
	\end{array}
	\end{equation*}
	The discrete transition relation is defined $\forall t_i \in \pnTransitions$ as:
	\begin{equation*}
	\begin{array}{l}
	(M, I) \stackrel{t_i}{\longrightarrow} (M', I') 
	\text{ iff } 
	\left\{ \begin{array}{l}  t_i \in enabled(M), \\
	M' = M - \pre{t_i} + \post{t_i}, \\
	\pnIntLeftPoint{I(t_i)} = 0,\\
	\forall t_k \in \pnTransitions,
	
	I'(t_k) =
	\left\{ 
		\begin{array}{l}
		I_s(t_k) \text{ if } t_k \in \uparrow enabled(t_k, M, t_i) \\
		I(t_k,) \;	\text{, otherwise }
		\end{array}
		\right. 
	\end{array} 
	\right.
	\end{array}
	\end{equation*}
\end{itemize}

\subsection{Parametric \tctl}

In this subsection, we begin by recalling the definition of \tpntctl 
\cite{boucheneb2009tctl}, that was inspired by \tctl 
\cite{alur1990model}
%\cite{alur_model-checking_1990} to tackle bounded time Petri nets. Then we
to tackle bounded time Petri nets. 
Then we give its parametric version introduced in \cite{Traonouez-JUCS-09}. 
These logics have been implemented 
in the \textsc{Rom{\'e}o} tool~\cite{lime2009romeo},
which allows to analyze timed extensions of Petri nets
 and perform parametric model-checking. 

But first, let us define \textit{Generalized Mutual Exclusion Constraints}, 
i.e. combinations of conjunctions and/or disjunctions of linear constraints 
that limit the weighted sum of tokens in a subset of places. 

\begin{definition}[\gmec\index{gmecs@\gmecs}]
  \label{def:gmecs}
  Let $\pnModel$ be a {\ptpn}.
  %(where $\pnPlaces=\{p_1,p_2,\ldots,p_n\}$ is the set of places). 
  A \gmec is inductively defined by:
  \begin{align*}
    \begin{split}
      \textit{\gmec} ::= & \left( \sum_{i=1}^{n} a_i
        *M(p_i)\right) \bowtie c \; | \; \varphi \vee \psi \;
      | \; \varphi \wedge \psi \; | \; \varphi \Rightarrow \psi
    \end{split}
  \end{align*}
  where $a_i \in \setZ$, $p_i \in \pnPlaces$,
  $\bowtie {}\in \set{<, \leq, =, >, \geq}$, $c \in \setN$ 
  and $\varphi,  \psi \in \gmec $, 
  the operators ($\vee$,$\wedge$,$\Rightarrow$) having their usual meaning.
\end{definition}

\begin{definition}[\tpntctl\index{logique temporelle!\tpntctl}]
\index{tpntctl@\tpntctl!definition}
  The temporal logics \tpntctl is inductively defined by: 
  \begin{align}
    \begin{split}
\varphi := \textsc{GMEC} \st \neg \varphi \st \varphi \Rightarrow \psi \st A\varphi\; {U}_{I} \psi  \st E\varphi\; {U}_{I} \psi 
    \end{split}
  \end{align}
  where $\textsc{GMEC}$ is a \gmec, $\varphi, \psi \in \textit{\tpntctl}$, 
  $I$ is an interval from 
  %$\setRp$ \todo{from $\N$?} with integer bounds s.t.
  $\N$ with integer bounds s.t.
$[n,m]$, $[n,m[$, $]n,m]$, $]n,m[$, or $[m,\infty[$, $n,m\in\setN$.
\end{definition}

% \medskip

The ($\neg$,$\imply$) operators have their classical meaning and we use the following aliases: $\mathbf{true}= \neg
\mathbf{false}$, $EF_{I} \phi = \exists\mathbf{true}\;
{\cal U}_{I} \phi$, $AF_{I} \phi = A\mathbf{true}\;
{\cal U}_{I} \phi$, $EG_{I} \phi = \neg AF_{I}
\neg \phi$, $AG_{I} \phi = \neg EF_{I} \neg
\phi$.

This logics can be parametrized the following way: 
\begin{definition}[\ptpntctl\index{logique temporelle par!\tpntctl}]
\index{tpntctlpar@\ptpntctl!definition}
  The parametric temporal logics \ptpntctl is inductively defined by: 
  \begin{align}
    \begin{split}
\varphi := E\varphi\;{U}_{I} \psi \, |\, A\varphi\; {U}_{I} \psi \, | \, E F_{I} \varphi \, | \, A F_{I} \varphi \, | \, E G_{I} \varphi \, | \, A G_{I} \varphi \, | \, \varphi \leadsto_{I_r} \psi
    \end{split}
  \end{align}
  where $\varphi$ and $\psi$ are \gmec, $I$ and $I_r$ are
  %\textbf{parametric intervals with integer bounds} 
  {parametric intervals with integer bounds} 
  s.t. $[n,m]$, $[n,m[$, $]n,m]$, $]n,m[$, or $[m,\infty[$, $n,m\in\setN$, with the restriction that $I_r=[0,m]$ or $I_r=[0,\infty[$. 
\end{definition}

Here, the bounded time response operator $\leadsto_{I_r}$ is 
defined as 
$A F \left( \varphi \imply A F_{I_r} \psi \right)$.

%\todo{$\leadsto$ has to be explained}

%In addition to this logical framework, 
%it is possible to define and use observers 
%%to model-check additional properties in TPNs and \ptpn\hspace{-0.5em}s.
%to model-check additional properties in TPNs and P-TPNs.
%%\todo{address "our" PRITPN here?}
%It consists in adding to the Petri net - in a non-intrusive manner - 
%\todo{more details are needed on the non-intrusive manner}
%places and transitions to model the property to check.
%\change{The property is encoded as a marking on the extended Petri net
%and we check its reachability~\cite{toussaint-FTDCS-97}.}
%\sout{The property is transformed in testing for the reachability of a given marking.}
%
%The main drawbacks of observers are two folds: 
%first, there is no automatic procedure to build them; 
%second, the observer can dramatically increase the size of 
%the state space to be explored by the model-checking procedure. 
%\todo{RV1: discuss the scalability of the approach here}
%%\subsection{Romeo tool}
%	We use the tool
%	\textsc{Rom{\'e}o}~\cite{lime2009romeo}\footnote{http://romeo.rts-software.org/}
%	that has been developed for the analysis
%	of parametric TPN (state space computation 
%	and "on the fly" model-checking of reachability properties 
%	and TCTL properties).
\section{Resilience properties of biological oscillatory systems}
\label{app:resilience_queries_section}
% % % REVIEW

\subsection{Resilience related \ptpntctl query specification}
\label{app:resilience_queries}
Properties~A-F
introduced in the main body of the paper
describe a certain set of behaviors that is normally 
exposed by the circadian clock model $\ccpn$.
%\sout{However we can enrich observers and modify the model itself 
%in such a way that allows to study the limits of its applicability
%by including unknown parameters 
%in the transition firing interval function.}
However we can study the applicability of the model
using the parameters 
in the transition firing interval function.
%	\paragraph{Robustness to the duration of day and night}
	The main 
	%\sout{Zeitgeber} 
	external stress
	in the framework of mammalian circadian clock
	is light (sunlight or artificial light).
	The distortion of the normal day-night cycle affects the nominal
	behavior 
%	\sout{of the circadian clock} 
	which causes negative effects
	like jet-lag.
	Let us consider how we can model the change of light conditions
	in the framework of our model.
	
	\begin{Query}
	\label{prop:cc_in_absence_of_light}
	Does property 
	$\phi_{\ref{prop:cc_in_absence_of_light}} = 
	\phi_{A}
	\wedge
	\phi_{B}
	\wedge
	\phi_{C}
	$
	%\todo{not formal enough}
	holds when light is always off?
	\end{Query}
	It is known~\cite{oster_disruption_2002} that the circadian 
	clock functions with a period of approximately $24$ hours 
	in the absence of light.
	In order to check this property we use the different 
	initial state which is also consistent with~\cite{comet_simplified_2012},
	namely $M(P_{\Lzero})=1$, $M(P_{\Gone})=0$ and $M(P_{\PCzero})=1$.
	We add an observer $O$ that prevents the light from changing its state,
	$O = \lbrace p_O \rbrace$, $M(p_O)=1$ 
	and $\inh{t_{on}} = p_O$.
	The property is satisfied by the model $\ccpn$
	with the values of parameters 
	%$\tau_{0,1}, \ldots$
	different from 
	those defined under the normal light 
	entrainment.
	
	In order to prevent the certain behavior in the system
	associated with transitions $T' \subset \pnTransitions$
	we add observers
	$O_t = \lbrace P_{O_t} \rbrace$, $M(P_{O_t})=1$
	for each $t \in T'$
	such that $\inh{t} = P_{O_t}$.
%	\todo{we don't need to add old inhibitor places since it is already
%	inhibited}

	Let us now enrich observers with parameters when it is possible.
	With that, we can check the limits of robustness 
	of a certain
	system under the perturbed environmental conditions.
	\begin{Query}
	\label{prop:cc_light_duration}
	What is the possible duration 
%	\sout{of the day} 
	of the period with light
	such that
	$\phi_{\ref{prop:cc_in_absence_of_light}}$ holds?
	\end{Query}
	In order to check this property, we add the observer that
	substitutes the original transition responsible for 
	switching the light off by another transition $t_*$
	with the parametric firing interval
	$\pnFiringIntervalFunction(t_*) = [\tau_d, \tau_d]$
	such that
	$O = \lbrace p_O, t_* \rbrace$, $M(p_O) = 1$,
	$\inh{t_{off}} = p_O$, 
	$\pre{t}_* = p_{\Lone}$ and $\post{t}_* = p_{\Lzero}$.
	We also have fixed the values of $\tau_{0,1}$ and
	$\tau_{1,0}$ 
	and $\tau_g = 1$
	%so that	the behavior of the original model is reflected
	to mimic the nominal behavior
	therefore limiting the possible search space for $\tau_d$.
	This property is satisfied by the model $\ccpn$, $\tau_g = 1$
	with $\tau_d \in [6,12]$.

	The last property we consider addresses the perturbation
	of the light conditions by switching on the light 
	during ''night''
	(the period with $M_{P_{\Lzero}}=1$
	that preserves the nominal behavior).
	
\begin{figure}
\centering
\includegraphics[width=0.6\linewidth]{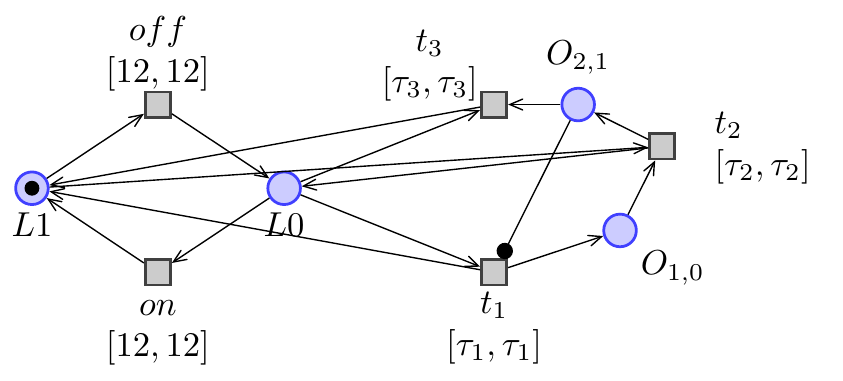}
\caption{Light switch observer during ''night''.}
\label{fig:nightswitch}
\end{figure}

	\begin{Query}
	\label{prop:light_during_night}
	For how long can the light be switched on during ''night''
	such that
	$\phi_{\ref{prop:cc_in_absence_of_light}}$ holds?
	\end{Query}
	In order to check this property, we  
	add the observer $O_1$ that inhibits the transition $t_{on}$
	and the observer $O_2$ that models switching 
	the light on during ''night''.
	We show the relevant part of the Petri in Figure~\ref{fig:nightswitch}
	(the detailed description of observers is omitted for the sake of readability).
	This property is satisfied by the model $\ccpn$,
	with $\tau_g - \tau_2 \geq 1$, $\tau_2 + \tau_3 \in [0,4]$,
	where we require $\tau_1 + \tau_2 + \tau_3  = 12$.
	Unfortunately it does not directly answer the stated question
	and more biologically inspired properties are needed
	to get more precise parameter estimations.

\subsection{Model elicitation}
We have formulated a series of properties as shown above.
By checking them, biologists may gain new inspirations about the
underlying biological processes
as well as it can be used to make the elicitation 
of the model that describes
certain biological phenomenon.
Here we provide the two example for circadian clock model.

\paragraph{Firing delay of transition $t_g$.}
When introducing the model in Figure~2, %\ref{fig:cc_pn},
we add only one constraint 
$\gamma = \lbrace \tau_g \geq 1 \rbrace$
so that it is not instantaneous.
We check the property
$\phi_{\ref{prop:cc_in_absence_of_light}}$,
where we fix the values of oscillation parameters,
namely 
$\tau_{0,1} = 18$,
$\tau_{1,0} = 6$
(property~A),
$\tau_{0,1} = 6$,
$\tau_{1,0} = 18$
(property~B),
and
$\tau_{0,1} = 5$
(property~C).
However, the parameter synthesis does not give any
additional information about the transition $t_g$.
We construct the observer from 
property~E 
for the transition $t_g$
and check the property $EF_{[0,\infty]} \left( M(p_{O,t_g}) > 0 \right)$,
and the latter does not hold.
Indeed, the only possible value of the parameter
$\tau_g = 0$ is biologically irrelevant.
It raises the question about the behaviors where transition $t_g$
is needed: it might be relevant only for a certain 
perturbation of environmental conditions.
We modify the model by making the firing intervals 
of transitions $off$ and $on$ parametric,
such that
$\pnFiringIntervalFunction(t_{off}) = [\tau_{off},\tau_{off}]$
and
$\pnFiringIntervalFunction(t_{on}) = [\tau_{on},\tau_{on}]$,
and inducing the additional condition on parameters
%$\gamma' = \lbrace \tau_{on} + \tau_{off} = 24  \rbrace$.
$\tau_{on} + \tau_{off} = 24$.
The property is then satisfied for 
$\tau_g \geq 1$, $\tau_{on} \in [7,11]$ and
$\tau_g \geq 1$, $\tau_{on} = 23$.
This shows that the model provided in~\cite{comet_simplified_2012}
initially allows for various biologically relevant behaviors.
%\sout{But the exact conditions under which such behaviors are observed
%can be difficult to find and \ptpntctl allows to address them.}

\paragraph{Firing delay of transition $t_a$.}
The model checking of property~E
shows that it is only satisfied with 
the zero firing delay $\tau_a = 0$ of the transition $t_a$.
As in the previous example, we may ask the question
about the environmental light conditions that allow 
for the firing of transition $t_a$ 
with the initial firing delay $\tau_a = 7$.
%\sout{Considering the same modified model as before, }
We check the property
$EF_{[0,\infty]} \left( M(p_{O,t_a}) > 0 \right)$, 
where the observer for transition $t_a$
is constructed in the same fashion 
as in property~E.
It is then satisfied with
$\tau_{on} \in \lbrace 23,24 \rbrace$.
The case $\tau_{on} = 24$ corresponds to 
property~I
where circadian clock in constant darkness is considered.
%\sout{This addresses the biological insight behind the
%transition $t_a$.}
We see that the set of admissible behaviors
such that $t_a$ is eventually fired is larger
than we predicted.
It shows as well that the model in~\cite{comet_simplified_2012}
has a certain redundancy that allows to 
address the change of environmental conditions.
Having enough biologically relevant knowledge it may
be possible to define the delays of all transitions
in the system using this approach.

This work-flow can be applied to any \ptpn model
that describes gene regulatory network.
%\todo{any limitations here?}
Having enough biological knowledge that can be 
formalized in terms of \ptpntctl properties,
such models can be refined 
together with the limits of their applicability.
%can be understood..
Model checking procedures can not address the 
pure inference problem,
therefore this preliminary knowledge is needed.

\subsection{Effects of gene knock-out and artificial jet-lag}
\label{app:gene_knock_out}
	The observers introduced above allow to study
	the behavior of the system after the gene knock-out
	and under the effect of artificial jet-lag.
	The effect of gene knock-out can be modeled by 
	simply suppressing all the behaviors that allow 
	the set of genes
	$\G$ to change its state from $0$ to $1$, that is
	by suppressing transitions $t_b$ and $t_f$.
	The behavior of the system $\ccpn$ is then restricted
	only to changes of light state and there is no
	permanent oscillations of protein $\PC$
	which means that gene knock-out leads to the
	system malfunction.
	
\begin{figure}
	\centering
	\includegraphics[width=0.5\linewidth]{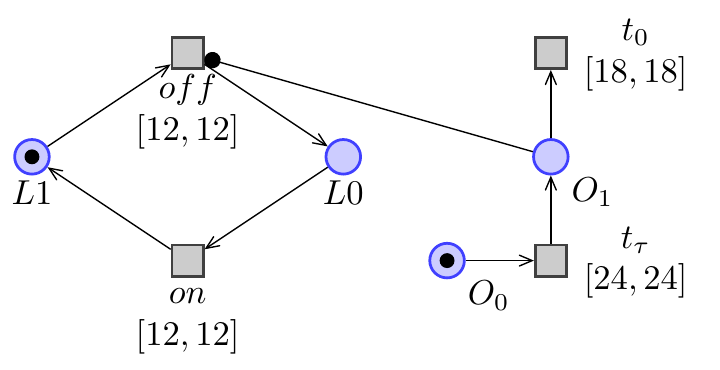}
	\caption{Artificial jet-lag observer}
	\label{fig:simplejetlag}
	\end{figure}

	Let us discover the effect of artificial jet-lag  on the model,
	when the duration of the period with light is prolonged
	(since the period without light does not affect the standard
	behavior much).
	We assume that there are no perturbations
	during first $24$ time units,
	%of the
	%light conditions 
	then the light is switched on for $30$ time units
	and
	after that the system returns to the nominal behavior.
	This effect can be modeled in a similar way 
	to query~\ref{prop:cc_in_absence_of_light}.
	We add the observer that prevents the light from switching off
	after $24$ time units and allows it again after $16$ time units.
%	\sout{,
%	therefore making the transition $t_{off}$ active again}.
	In this way it is guaranteed that $M(p_{\Lone})=1$ for $30$ time units
	without any switching.
	We show the relevant part of the Petri net
	with the observer in Figure~\ref{fig:simplejetlag}.
	Checking the property~A
	shows that the system does not function normally
	as the time for $\PC$ to change the state 
	from $0$ to $1$ is $\tau_{0,1} \geq 36$.
	This corresponds to the fact that $t_c$ is the only transition
	that changes the state of $\PC$ from $0$ to $1$ 
	and it is enabled only when light is switched off.
	It is important to notice that \ptpn formalism 
	does not allow to model the recovery process of the system
	since it is ruled by local clocks only.
	For instance, if the delay observer $'O$ is added
	with a delay of $100$ time units,
	the result of model checking the property~A
	after this delay
	refers to the standard behavior, i.e. $\tau_{0,1} \geq 18$.	

\subsection{Resilience in \ptpntctl formalism}
Here we stated the number of properties that describe the
standard behavior of the circadian clock model
such as permanent oscillation
and entrainment properties of the set of genes $\G$.
Please note that we can judge only about certain local time properties
meaning that there is no notion of the global clock in 
\ptpn models.
Due to the limited expressivity of \ptpntctl, we enrich
the model with observers that are given in a fairly
general fashion.
We also addressed the limits of the model applicability
by enriching the observers with parameters that are
determined by parameter synthesis procedure.
It allows us to verify resilience related properties
such as robustness to changes in the environmental
conditions.
%The latter is also extended by considering the
%influence of artificial jet-lag and gene knock-out.

%since the property 
%$\phi_{\ref{prop:cc_in_absence_of_light}}$
%holds under the same restriction 
%$\gamma = \lbrace \tau_g \geq 1 \rbrace$.

%	\phi_{\ref{prop:pc_always_oscillation}}
%	\wedge
%	\phi_{\ref{prop:g_always_oscillation}}
%	\wedge
%	\phi_{\ref{prop:g_entrain_by_g_and_p}}

%\begin{itemize}
%	\item Resilience intuition and definition
%	\item "Resilience", "reactive" and "adaptive"
%	\item Model consistency
%		Every possible model of biological system shall satisfy 
%		certain consistency specification.
%		Here we require that there are no redundant transitions
%		in the system. \sr{formal definition is required}
%	\item Entrainment
%	\item Coupling
%	\item External stress effect
%	\item Formulate what we mean under nominal behaviour
%	\item Finally formulate resilience of the biological oscillatory system
%\end{itemize}

%\sr{Since we are considering only Boolean variables, there is
%no notion of the amplitude of the oscillation.\todo{But you told me you were considering }}
\end{appendix}

\bibliographystyle{splncs03}
\bibliography{pn_bibfile}

\begin{thebibliography}{1}
\providecommand{\url}[1]{\texttt{#1}}
\providecommand{\urlprefix}{URL }

\bibitem{alur1990model}
Alur, R., Courcoubetis, C., Dill, D.: Model-checking for real-time systems. In:
  Logic in Computer Science, 1990. LICS'90, Proceedings., Fifth Annual IEEE
  Symposium on Logic in Computer Science. pp. 414--425. IEEE (1990)

\bibitem{tpn_expressive_power_2013}
B{\'e}rard, B., Cassez, F., Haddad, S., Lime, D., Roux, O.H.: The expressive
  power of time petri nets. Theoretical Computer Science  474,  1--20 (2013)

\bibitem{boucheneb2009tctl}
Boucheneb, H., Gardey, G., Roux, O.H.: {TCTL} model checking of time petri
  nets. Journal of Logic and Computation  19(6),  1509--1540 (2009)

\bibitem{comet_simplified_2012}
Comet, J.P., Bernot, G., Das, A., Diener, F., Massot, C., Cessieux, A.:
  Simplified models for the mammalian circadian clock. Procedia Computer
  Science  11,  127--138 (2012)

\bibitem{larsen1995model}
Larsen, K.G., Pettersson, P., Yi, W.: Model-checking for real-time systems. In:
  Fundamentals of computation theory, pp. 62--88. Springer (1995)

\bibitem{lime2009romeo}
Lime, D., Roux, O.H., Seidner, C., Traonouez, L.M.: Romeo: A parametric
  model-checker for {P}etri nets with stopwatches. In: Tools and Algorithms for
  the Construction and Analysis of Systems, pp. 54--57. Springer (2009)

\bibitem{oster_disruption_2002}
Oster, H., Yasui, A., van~der Horst, G.T.J., Albrecht, U.: Disruption of
  {mCry}2 restores circadian rhythmicity in {mPer}2 mutant mice. Genes \&
  Development  16(20),  2633--2638 (2002)

\bibitem{Traonouez-JUCS-09}
Traonouez, L.M., Lime, D., Roux, O.H.: Parametric model-checking of stopwatch
  {P}etri nets. Journal of Universal Computer Science  15(17),  3273--3304
  (2009)

\end{thebibliography}

\end{document}